\documentclass[aps,prc,twocolumn,showpacs,groupedaddress,floatfix]{revtex4}

\usepackage{graphicx}
\usepackage{dcolumn}
\usepackage{bm}

\def\BM#1{\mbox{$\displaystyle B^M_{#1}$}}

\begin{document}

\title{Coordinate Space HFB Calculations for the\\
       Zirconium Isotope Chain up to the Two-Neutron Dripline}
\author{A. Blazkiewicz, V.E. Oberacker,  A.S. Umar}
\affiliation{Department of Physics and Astronomy, Vanderbilt University,
     Nashville, Tennessee 37235, USA }

\author{M. Stoitsov}
\affiliation{Department of Physics and Astronomy,
University of Tennessee, Knoxville, TN 37996}
\affiliation{Physics Division, Oak Ridge National
 Laboratory, P.O. Box 2008, Oak Ridge, TN 37831}
\affiliation{Joint Institute for Heavy-Ion Research, Oak Ridge, TN 37831}

\affiliation{Institute of Nuclear Research and Nuclear
 Energy, Bulgarian Academy of Science, Sofia 1784, Bulgaria}
\date{\today}


\begin{abstract}
We solve the Hartree-Fock-Bogoliubov (HFB) equations for deformed,
axially symmetric even-even nuclei in coordinate space on a 2-D
lattice utilizing the Basis-Spline expansion method. Results are
presented for the neutron-rich zirconium isotopes up to the
two-neutron dripline. In particular, we calculate binding energies,
two-neutron separation energies, normal densities and pairing
densities, mean square radii, quadrupole moments, and pairing gaps.
Very large prolate quadrupole deformations
($\beta_2=0.42,0.43,0.47$) are found for the $^{102,104,112}$Zr
isotopes, in agreement with recent  experimental data. We compare
2-D Basis-Spline lattice results with the results from a 2-D HFB code which
uses a transformed harmonic oscillator basis.
\end{abstract}
\maketitle


\section{\label{sec:intro}Introduction}
In contrast  to the well-understood  behavior near the valley of stability, there
are many open questions  as we move towards the neutron dripline.
In this exotic region of the  nuclear chart, one expects to see several new
phenomena \cite{NNigata}: the neutron-matter distribution is very diffuse
giving rise to the ``neutron skins'' and the ``neutron halos''. One also expects
new collective modes associated with the neutron skin, e.g. the ``scissors''
vibrational mode.

In current experiments, the neutron dripline has only been reached for relatively
light nuclei. The second-generation radioactive ion beam (RIB) facilities currently
under construction will allow us to explore nuclear structure and astrophysics
at the driplines of heavier nuclei.

It is generally acknowledged  that an accurate theoretical treatment of the
pairing interaction
is  essential for a description of  the  exotic nuclei \cite{TOU03,OU03}.
Besides large  pairing correlations, the HFB calculations have to
face the  problem of an  accurate description of the continuum states with a
large spatial extent. All of these  features represent major challenges for
the numerical solution.

There are various  mean-field methods  of solving the
non-relativistic HFB equations. Generally, they can be divided into
two categories: lattice methods and basis expansion methods. In the
lattice approach no region of  the spatial lattice is favored over
any other region: the well bound, weakly bound and (discretized)
continuum states can be represented with the same accuracy. Currently, there
are only two codes which solve the HFB problem on a lattice in
coordinate space without any approximations using the
quasiparticles: a 1-D (spherical) HFB code \cite{DN96} and
the present 2-D (axially symmetric) code \cite{TOU03,OU03,OU04}. In
the basis expansion method a wave function is expanded into the
chosen basis functions: the harmonic oscillator basis  (HO)
\cite{egidorobledo}, the transformed harmonic oscillator basis (THO)
\cite{Stoitsov:2003pd}, the HF orbitals
\cite{yamagami,Terasaki:1995du}. There is also an alternative method
that is under development where the HFB equations are solved in the
3-D Cartesian mesh using the canonical-basis approach \cite{Tajima2003}.

We solve the Hartree-Fock-Bogoliubov (HFB) equations for deformed, axially symmetric nuclei
in coordinate space on a 2-D lattice \cite{TOU03,OU03}. Our computational technique
(the Basis-Spline collocation and Galerkin method) is  particularly well suited to
study the ground state properties of nuclei near the driplines. It allows us to take into
account high-energy continuum states up to an equivalent single-particle energy of 60 MeV
or more.

In this paper we  study the ground state properties of
neutron-rich zirconium nuclei ($Z=40$) up to the two-neutron
dripline. The isotope chain calculations start from the $^{102}$Zr
isotope up to the dripline nucleus, which turns out to be
$^{122}$Zr. For a long time, the $A \sim 100$ region  has been  of interest
to nuclear structure physicists as an area of competition between
various coexisting nuclear shapes (well-deformed prolate, oblate,
or spherical) \cite{Ca02}. The zirconium isotopes are known to
possess a rapidly changing  nuclear  shape when the neutron number
changes from 56 to 60 \cite{R-A04}. We find that
a spherical ground state shape is preferred over a
prolate shape starting from the $^{114}$Zr isotope up to the dripline nucleus
$^{122}$Zr. In this paper, we compare  Basis-Spline lattice results with  the HFB-2D-THO code
\cite{Stoitsov:2003pd}, which  uses a transformed harmonic
oscillator basis (THO), as well as with currently available experimental data.


\section{\label{sec:hfb_eqns}Skyrme-HFB equations in coordinate space}

A detailed description of the Basis-Spline lattice method has been published in
references \cite{TOU03,OU03}; we give here only a brief summary.
In coordinate space representation, the HFB Hamiltonian and the quasiparticle
wavefunctions depend on the distance vector ${\bf r}$, spin projection
$\sigma = \pm \frac{1}{2}$, and isospin projection $q = \pm \frac{1}{2}$
(corresponding to protons and neutrons, respectively).
In the HFB formalism, there are two types of quasiparticle (bi-spinor) wavefunctions,
$\phi_1$ and $\phi_2 $.
In the present work, we utilize a Skyrme effective N-N interaction in the particle-hole
channel while the $p-p$ and $h-h$ channels are described by a zero-range delta pairing
force. For these types of effective interactions, the particle mean field Hamiltonian
$h$ and the pairing field Hamiltonian $\tilde h$ are diagonal in isospin space and
local in position space, and the HFB equations have the following structure \cite{TOU03}:
\begin{equation}
\left(
\begin{array}{cc}
( h^q -\lambda ) & \tilde h^q \\
\tilde h^q & - ( h^q -\lambda )
\end{array}
\right)
\left(
\begin{array}{c}
\phi^q_{1,\alpha} \\
\phi^q_{2,\alpha}
\end{array}
\right)
 = E_\alpha
\left(
\begin{array}{c}
\phi^q_{1,\alpha} \\
\phi^q_{2,\alpha}
\end{array}
\right) , \label{eq:hfbeq2}
\end{equation}
where the Lagrange parameter $\lambda $ is introduced to yield the
correct particle-number on average since the HFB wavefunction does
not have a well defined particle-number.

The quasiparticle energy spectrum is discrete for $|E|<-\lambda$
and continuous for $|E|>-\lambda$ \cite{DN96}. For even-even nuclei it is customary to
solve the HFB equations for positive quasiparticle energies and consider all negative
energy states as occupied in the HFB ground state.
The quasiparticle wavefunctions determine the normal density
$\rho_q({\bf r})$ and the pairing density $\tilde \rho_q({\bf r})$
as follows
\begin{eqnarray}
\rho_q({\bf r}) \ &=& \ \sum_{E_\alpha > 0}^{\infty} \sum_{\sigma = -\frac{1}{2}}^{+\frac{1}{2}}
       \phi^q_{2,\alpha} ({\bf r} \sigma) \ \phi^{q \ *}_{2,\alpha} ({\bf r} \sigma) \ ,
\label{eq:density}    \\
\tilde{\rho_q}({\bf r}) \ &=& \ - \sum_{E_\alpha > 0}^{\infty} \sum_{\sigma = -\frac{1}{2}}^{+\frac{1}{2}}
       \phi^q_{2,\alpha} ({\bf r} \sigma) \ \phi^{q \ *}_{1,\alpha} ({\bf r} \sigma) \ .
\label{eq:pairing_density}
\end{eqnarray}

Restricting ourselves to axially symmetric nuclei, we use cylindrical coordinates
$(r,z,\phi)$. In the HFB lattice method, we introduce a 2-D grid
$(r_\alpha,z_\beta)$ with $\alpha = 1,...,N_r$ and $\beta = 1,...,N_z$.
In radial direction, the grid spans the region from $0$ to $r_{max}$.
Because we want to be able to treat octupole shapes, we do not assume left-right
symmetry in $z$-direction. Consequently, the grid extends from $-z_{max}$ to
$+z_{max}$. Typically, $z_{max} \approx r_{max}$ and $N_z$ $\approx$ $2 \cdot N_r$.
The HFB wavefunctions and operators are represented on the 2-D lattice by
Basis-Spline expansion techniques. In practice, B-Splines of order M=9 are
being used. Details about the B-Spline collocation and Galerkin
methods are given in the Appendix.


\section{Numerical results and comparison with experimental data}
This section describes the numerical results for the even-even
members of the zirconium ($Z=40$) isotope chain up to the
two-neutron dripline. Two theoretical approaches are presented and
compared: a) a 2-D lattice method using B-Spline technology
(hereafter referred to as HFB-2D-LATTICE), and b) an expansion in a
2-D harmonic oscillator (HFB-2D-HO) and transformed harmonic
oscillator basis (HFB-2D-THO). We also compare  theoretical
results with experimental data whenever possible.

In all calculations we utilize the SLy4 Skyrme force \cite{CB98} and
a zero-range pairing force with a strength parameter $V_0 =-187.1305$
(HFB-2D-LATTICE code)
and an equivalent single particle energy cutoff parameter of
$\varepsilon_{max}=60$ MeV.
The pairing strength V$_0$ has been
adjusted in both codes to reproduce the measured average neutron
paring gap of $1.245$ MeV in $^{120}$Sn \cite{DN96}, as can be seen
from Table \ref{table:Sn120}.
\begin{table}[hbt!]
\caption{\label{table:Sn120}
Results for $^{120}_{\ 50}Sn$. Comparison of results obtained from the HFB-2D-LATTICE
code (first row) with HFB-2D-THO results (second row) and the experiment
(third row). The columns display binding energies (BE), intrinsic quadrupole
moments for neutrons and protons (Q$_n$,Q$_p$), rms-radii (r$_n$,r$_p$),
average pairing gaps ($\Delta_n,\Delta_p$),
pairing energy for neutrons (PE$_n$), and Fermi levels ($\lambda_n,\lambda_p$).
}
\begin{ruledtabular}
\begin{tabular}{ l c c c l l  }
             & BE(MeV)    & Q$_n$(fm$^2$)  & Q$_p$(fm$^2$)  & r$_n$(fm) & r$_p$(fm)  \\
\hline
             & -1019.26    &  0.29      & 0.12           &4.725 &  4.590 \\
             & -1018.22    &  0.00      & 0.00           &4.728   & 4.593     \\
             & -1020.54    &   -            &-               & -  &-    \\
\hline
             & $\Delta_n$(MeV)  & $\Delta_p$(MeV)      & PE$_n$(MeV)  & $\lambda_n$(MeV) & $\lambda_p$(MeV) \\
 \hline
             & 1.244999         & 0.0                  &-10.24       &-7.98              &-8.16   \\
             & 1.245469         & 0.0                  &-10.26       &-7.99              &-11.13 \\
             & 1.245            & -                    & -           & -  &-
\end{tabular}
\end{ruledtabular}
\end{table}

In  Table \ref{table:Sn120} we compare the results from the two HFB
codes for the $^{120}_{\ 50}$Sn isotope. All observables agree very
well. The apparent ``disagreement'' in the proton Fermi level
$\lambda_p$ is really an artifact: the pairing gap vanishes at the
magic proton number $Z=50$ resulting in an ill-defined Fermi energy.
The two codes use different prescriptions for calculating
$\lambda_p$ in the trivial case of no pairing.  HFB+THO accepts
the last occupied equivalent single-particle energy as $\lambda_p$
in this no pairing case, whereas the Basis-Spline code takes the
average of the last occupied and first unoccupied equivalent single-particle
energy levels.

\subsection{Deformations, dripline and  pairing properties}
Recently, triple-gamma coincidence experiments have been carried out with Gammasphere
at LBNL \cite{HR03} which have determined half-lives and quadrupole deformations of
the neutron-rich  $^{102,104}Zr$ isotopes. The isotopes from that region are produced in
the process of fission of transuranic elements and have been studied
via $\gamma$-ray spectroscopy  techniques.
These medium-mass nuclei are among the most neutron-rich isotopes
($N/Z \approx 1.6$) for which spectroscopic data are available.
Very large prolate quadrupole deformations
($\beta_2=0.43,0.45$) are found for the $^{102,104}$Zr isotopes. Furthermore,
the laser spectroscopy measurements \cite{Ca02} for the zirconium isotopes have yielded
precise rms-radii in this region. Recently, an experiment has been carried out 
to measure the mass of the $^{104}$Zr isotope \cite{R-A04}.
It is therefore of a great interest to compare these data with the predictions of the
self-consistent HFB mean field theory.

\begin{figure}[!htb]
\includegraphics*[scale=0.35]{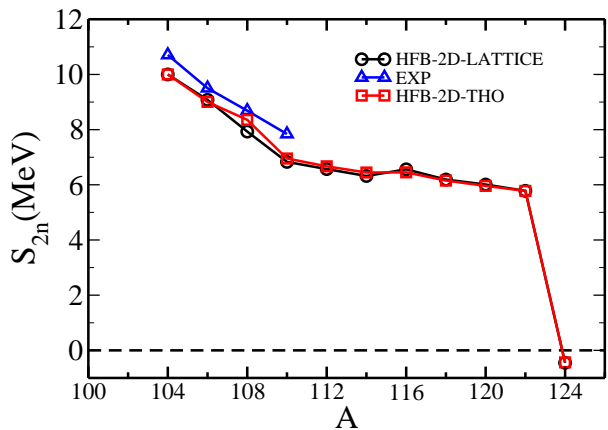}
\caption{\label{fig:fig1} (Color online) Two-neutron separation energies for
the neutron-rich zirconium isotopes. The dripline is located where the separation
energy becomes zero. The $^{122}$Zr isotope  is the last stable nucleus against two
neutron  emission.}
\end{figure}

In radial ($r$) direction, the lattice extends from $0 - 15$~fm, and in symmetry axis ($z$)
direction from $-15,...,+15$~fm, with a lattice spacing of about $0.8$~fm in the central
region. Angular momentum projections $\Omega = 1/2, 3/2, ..., 21/2$ were taken into account.
Figure~\ref{fig:fig1} shows the calculated two-neutron separation energies
for the zirconium isotope chain. The two-neutron separation energy is defined as
\begin{equation}
S_{2n}(Z,N) \ = \ E_{bind} (Z,N) \ - \ E_{bind} (Z, N - 2) \; .
\end{equation}
Note that in using this equation, all binding energies must be
entered with a \emph{positive} sign. The position of the two-neutron
dripline is defined by the condition $S_{2n}(Z,N) \ = 0 $, and
nuclei with negative two-neutron separation energy are unstable
against the emission of two neutrons. As one can see,
both codes (HFB-2D-THO  and HFB-2D-LATTICE) predict the dripline nucleus to be
at mass number $A= 122$. In addition, 
we observe a very good agreement   between the two codes in the
whole mass region of the isotope chain.
We also give a comparison with  the latest available experimental
data up to the  isotope  $^{110}$Zr  \cite{audi2003}.

\begin{figure}[!htb]
\includegraphics*[scale=0.35]{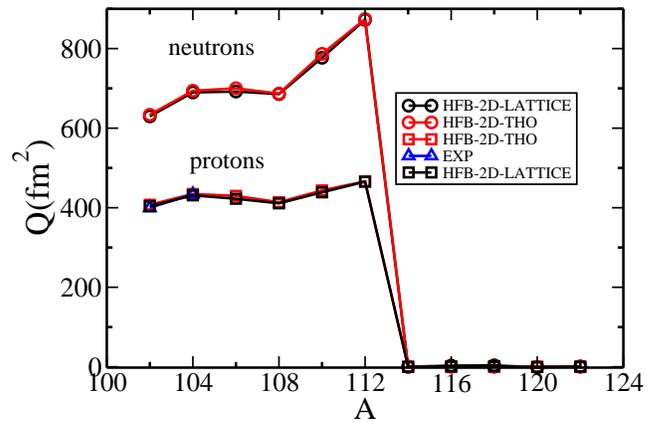}
\caption{\label{fig:fig2} (Color online) Intrinsic quadrupole moments for
protons and neutrons.}
\end{figure}

In Fig.~\ref{fig:fig2} we compare the intrinsic proton and neutron quadrupole moments
calculated with the lattice code and the THO code. Available experimental data \cite{HR03} are
also given.
Generally, we observe  a nearly perfect agreement between the two
codes as well as with the experiment.
The deformations (for neutrons ) in terms of the deformation
parameter $\beta_2$ for those nuclei,  namely for the $^{102-112}$Zr
isotopes   range from $\beta_2$=0.42 to $\beta_2=0.47$. Both the Basis-Spline lattice code
and the HFB-2D-THO code predict the $^{112}$Zr isotope to have the
largest ground state deformation. For mass numbers larger than
112, we observe spherical ground state shapes.
Experimental deformations for
protons  are available for  two isotopes, $^{102}$Zr and $^{104}$Zr
\cite{HR03}. Calculations agree with the experiment reasonably well
and give $\beta_2$ values of 0.42,0.43 while the experiment predicts
$\beta_2^{102}$=0.42, $\beta_2^{104}$=0.45.

\begin{figure}[!htb]
\includegraphics*[scale=0.33]{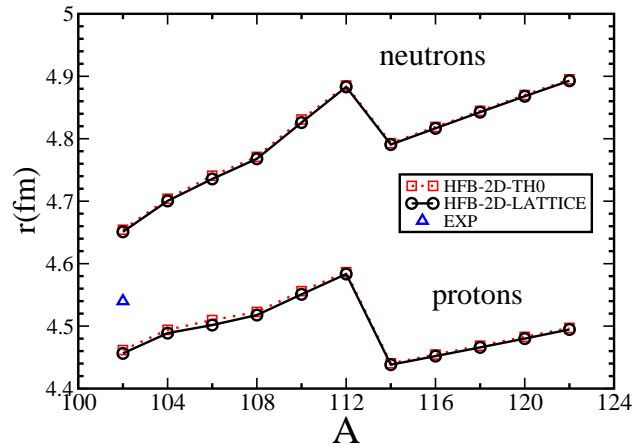}
\caption{\label{fig:fig3} (Color online) Root-mean-square radii for the chain of zirconium
isotopes. }
\end{figure}

In Fig.~\ref{fig:fig3} we compare the root-mean-square radii of protons
and neutrons predicted by the LATTICE code and the THO code.
Both codes give nearly identical results for the whole isotope chain.
Only one experimental data point is available, the proton rms radius of $^{102}$Zr
\cite{Ca02}. The experiment yields a proton rms radius of 4.54~fm
while the HFB codes predict a value of
4.45~fm (HFB-2D-LATTICE) and 4.46~fm (HFB-2D-THO). The difference between
theory and experiment is quite small, of order 2$\%$.
We can clearly observe the presence of the neutron-skin manifested by
the large differences between  the neutron and proton rms radii for
all of the isotopes in the chain. As expected the neutron-skin becomes
"thicker" as we approach the dripline. Starting at the mass number A=114
up to the  dripline the nuclei prefer a spherical ground state shape
(Fig.~\ref{fig:fig2}) which results in a sudden shrinking
of the rms radius  at A=114.

\begin{figure}[!htb]
\includegraphics*[scale=0.3]{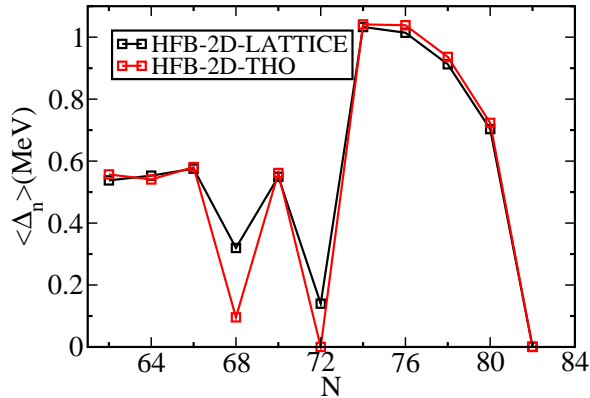}
\caption{\label{fig:fig4} (Color online) Average neutron pairing gap
 for the chain of zirconium isotopes. }
\end{figure}

\begin{figure}[!htb]
\includegraphics*[scale=0.3]{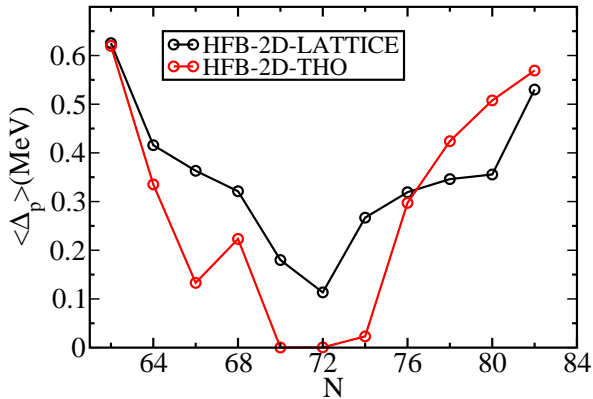}
\caption{\label{fig:fig5}  (Color online) Average proton pairing gap
for the chain of zirconium isotopes. }
\end{figure}

Fig.~\ref{fig:fig4} and Fig.~\ref{fig:fig5} depict the average
pairing gaps for neutrons and protons. Generally, both HFB codes
show the same trend for the pairing gaps as a function of neutron
number; the agreement is noticeably better for neutrons.
The two HFB codes  predict a
small value of the neutron pairing gap for the $^{112}$Zr isotope
which on the other hand has the largest prolate
deformation (Fig.~\ref{fig:fig2}) among the calculated nuclei.
Coincidentally, the dripline turns out to be at the neutron
magic number (N=82) and, as expected, both codes yield a
pairing gap of zero for the $^{122}$Zr isotope.
\begin{figure}[!htb]
\includegraphics*[scale=1.5]{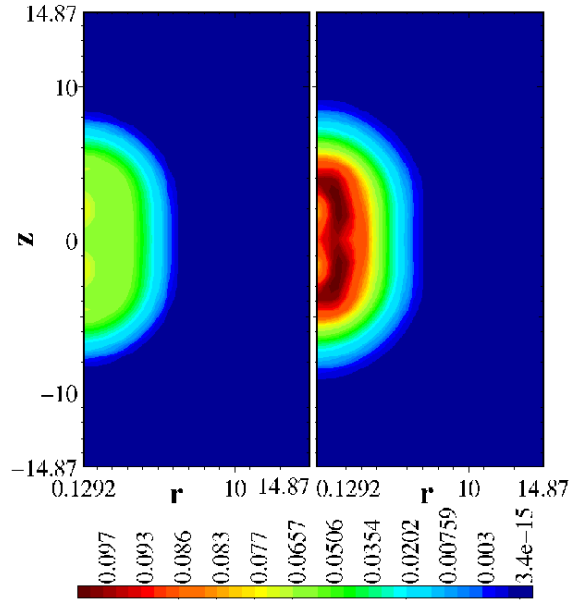}
\caption{\label{fig:fig6} (Color online)
Contour plots of the $^{110}$Zr normal densities, for protons
(left) and neutrons (right). Densities are shown as a function of
the cylindrical coordinates $(r,z)$, where $z$ is the symmetry axis.}
\end{figure}

\subsection{Density studies}

\begin{figure}[!htb]
\includegraphics*[scale=1.5]{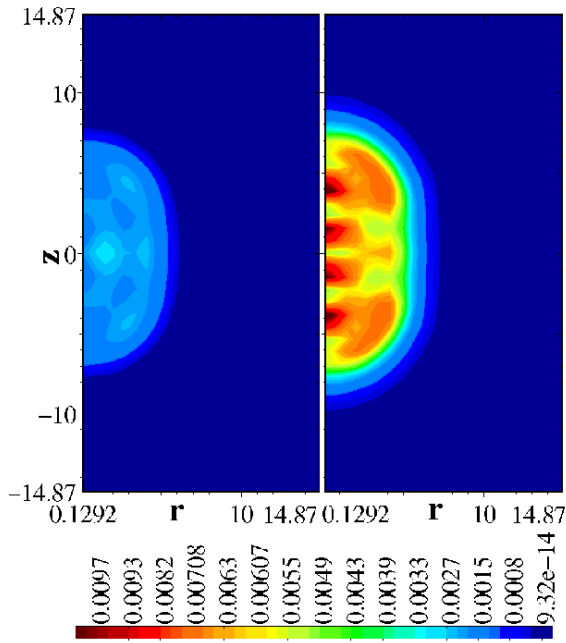}
\caption{\label{fig:fig7} (Color online)
Contour plots of the $^{110}$Zr pairing densities, for protons
(left) and neutrons (right). The densities are shown as a function of
the cylindrical coordinates $(r,z)$, where $z$ is the symmetry axis.}
\end{figure}

In this section we focus on the  normal and pairing
densities for the selected isotopes.
In Fig.~\ref{fig:fig6} we show a contour plot of normal densities
for protons and neutrons for the $^{110}$Zr isotope. It is the last deformed
isotope with a significant value of the pairing gap for neutrons,
(Fig.~\ref{fig:fig4}),
therefore it is possible to show plots of both normal and pairing densities.
The results obtained for the neutron normal and pairing
densities (Figs.~\ref{fig:fig6} and \ref{fig:fig7})
clearly exhibit a large prolate deformation. The normal density for neutrons
(Fig.~\ref{fig:fig6}) is concentrated in the region that extends from
$0$~fm to $2$~fm in $r$-direction and from $-5$~fm to $+5$~fm in $z$-direction.
Within this region, we find an enhancement in the neutron density with a
shape that resembles the figure ``eight''. In comparing the neutron to the
proton density, one notices that both the center
of the nucleus and  the surface is dominated by neutrons.

The pairing density for neutrons in Fig.~\ref{fig:fig7} shows
a richer structure  than the normal density. This quantity
describes the probability of correlated nucleon pair formation
with opposite spin projection, and it determines the pair transfer formfactor.
We can see that most  correlated  pair formation  takes place in the four
closed shaped structured areas near the z-axis. We may conclude that
neutrons dominate the pairing properties in the interior of this nucleus. The same
applies to normal densities (Fig.~\ref{fig:fig6}), yet the  difference
between  neutrons  and protons is more  pronounced
in case of the pairing densities. A graph depicting the single-particle energy spectrum
of the pairing density for the $^{104}Zr$ isotope has been published in Ref.~\cite{OU04}.

\begin{figure}[!htb]
\includegraphics*[scale=0.9]{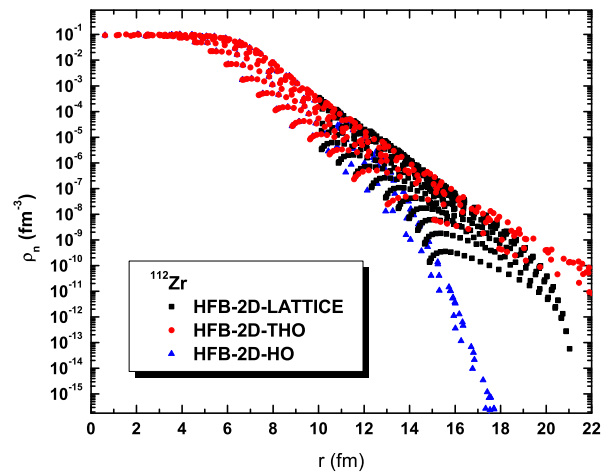}
\caption{\label{fig:fig8} (Color online)  Logarithmic plot
of the normal neutron density for the most
deformed isotope $^{112}$Zr as a function of the distance
$r={\sqrt{\rho^2+z^2}}$. }
\end{figure}

\begin{figure}[!htb]
\includegraphics*[scale=0.9]{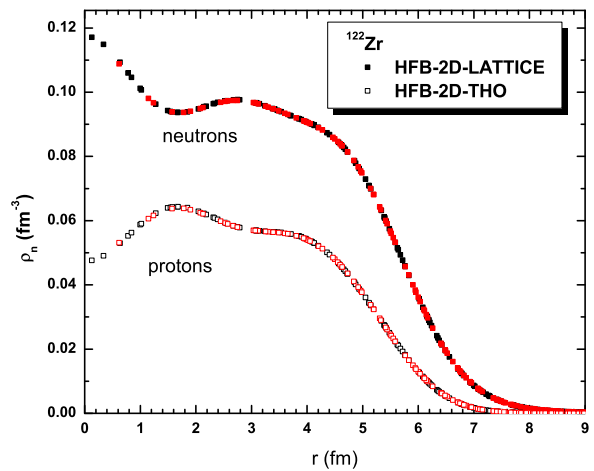}
\caption{\label{fig:fig9} (Color online) Linear plot of the
normal neutron and proton density for the dripline nucleus $^{122}$Zr as a
function of distance $r={\sqrt{\rho^2+z^2}}$. Comparison between the
HFB-2D-THO code and the HFB-2D-LATTICE code.}
\end{figure}

In Figs.~\ref{fig:fig8} and \ref{fig:fig9} we show plots of
normal densities as a function of the distance from the center,
$r={\sqrt{\rho^2+z^2}}$. For a given value of $r$, the density is
single-valued for a spherical nucleus and multi-valued for a
deformed density distribution because in the latter case different
combinations of lattice points $z_i$ and $\rho_j$ give rise to the
same $r$-value.

In Fig.~\ref{fig:fig8} we compare three different calculations of
the neutron normal density for the most deformed $^{112}$Zr isotope.
The plot on a logarithmic scale shows that the density distribution
predicted by the HFB-2D-THO and HFB-2D-LATTICE codes
is deformed for almost all values of the distance from the nuclear center,
$r$. At very large distances the densities become less
deformed since nuclear potentials go to zero and HFB equations lead
to a spherical asymptotic solution.
Fig.~\ref{fig:fig8} also shows for comparison the HFB-2D-HO result as an
illustration of the shortcomings of the pure harmonic oscillator basis
calculations to reproduce density distributions asymptoticly at very large
distances. One can see its too rapid decay beyond distances of about
12~fm. Clearly, the pure harmonic oscillator basis calculations
cannot represent properly density asymptotic for nuclei close to the
neutron drip line.

Neutron and proton normal densities for the drip-line nucleus
$^{122}$Zr are shown in  Fig.~\ref{fig:fig9}. From the single-valued plot
as a function of $r={\sqrt{\rho^2+z^2}}$ one can immediately conclude
that both neutron and proton normal densities are spherical. Another
striking feature is a strong neutron enhancement at the center and a
corresponding depletion in the proton density. Note also that the neutron
density is substantially larger than the proton density for all values
of $r$.


\section{\label{sec:conclusions}Conclusions}

In this paper, we have performed Skyrme-HFB calculations in coordinate space
for the neutron-rich zirconium isotopes up to the two-neutron dripline. We
have calculated the ground state properties  (even-even nuclei)  for the
zirconium isotopes (Z=40), starting from N=62 up to the
two-neutron dripline, which our HFB codes predict to be N=82. In particular,
we have calculated the two-neutron separation energies, quadrupole
moments, rms radii, average pairing gaps and densities.
In comparing HFB-2D-LATTICE theoretical calculations  for the two-neutron separation
energies (Fig.~\ref{fig:fig1}) with  the  HFB-2D-THO code we
find  the same  results. Particularly,  both codes
predict the $^{122}$Zr  isotope to be the dripline  nucleus.
We find very large prolate deformations for the $^{102-112}$Zr isotopes
and  a  spherical  ground state shape for the $^{114-122}$Zr nuclei.
The $\beta_2$ value  for the most deformed nucleus $^{112}$Zr in the
calculated chain  reaches an impressive value of 0.47.
The root-mean-square radii clearly show the existence of a
``neutron skin'' in the neutron-rich zirconium isotopes. We can also
observe a sudden shrinking of the rms-radius 
at A=114 due to a  change of the  prolate ground state deformation into
the  spherical shape. In section III.B. we studied
normal and pairing densities. In particular, Figures~\ref{fig:fig6} and \ref{fig:fig7}
show  the dominant nature of neutrons, where one can observe
that both normal and pairing densities are dominated by neutrons.
The density studies in Figure~\ref{fig:fig9}
predict the presence of a sizable neutron skin in the spherical dripline nucleus
$^{122}$Zr.


\appendix*

\section{Basis-Spline representation of wavefunctions and operators}

In this appendix, we discuss the representation of differential operators and
wavefunctions in terms of Basis-Splines which is used in the HFB-2D-LATTICE code.
There exists an extensive literature on Basis-Spline theory developed by mathematicians
\cite{DeB78}. We have adapted some of these methods for the numerical solution of
problems in atomic and nuclear physics on a lattice.  Ref.~\cite{Uma91}
discusses the B-Spline collocation method, periodic and fixed boundary
conditions, and the solution of various 1-D radial problems (Schr\"odinger, Poisson and
Helmholtz equations) and the solution of 3-D Cartesian problems (Poisson equation).
In a later paper, Umar et al.~\cite{Uma91a} solved the HF and TDHF equations on a
3-D lattice. In 1995, Wells et al.~\cite{WO95} applied the B-Spline collocation
method to the static and time-dependent Dirac equation which eliminated
the ``fermion doubling problem'' that one encounters with the finite-difference method.
In 1996, Kegley et al.~\cite{KO96} studied 2-D axially symmetric systems in cylindrical
coordinates with the collocation method and also introduced the Basis-Spline Galerkin
method. In the following, we compare both of these methods.

\subsection{HFB lattice representation}
For the lattice representation of wavefunctions and operators, we use a
Basis-Spline method.
Basis-Spline functions $B_i^M(x)$ are piecewise-continuous polynomials
of order $(M-1)$. They represent generalizations of finite elements
which are B-splines with $M=2$.

We consider an arbitrary (differential) operator equation
\begin{equation}
{\cal{O}} \bar{f}(x) - \bar{g}(x) = 0\;. \label{bspleq1}
\end{equation}
Special cases include eigenvalue equations of the HF/HFB type where ${\cal{O}}=h$
and $\bar{g}(x)=E \bar{f}(x)$.
We assume that both $\bar{f}(x)$ and $\bar{g}(x)$ are well approximated by spline
functions
\begin{eqnarray}
\label{bspleq2}
\bar{f}(x) \approx f(x) &\equiv& \sum_{i=1}^{\cal N}\BM{i}(x)a^i\; ,\nonumber \\
\bar{g}(x) \approx g(x) &\equiv& \sum_{i=1}^{\cal N}\BM{i}(x)b^i\;.
\end{eqnarray}
Because the functions $f(x)$ and $g(x)$ are approximations to the exact functions
$\bar{f}(x)$ and $\bar{g}(x)$, the operator equation will in general only be
approximately fulfilled
\begin{equation}
{\cal{O}} f(x) - g(x) = R(x)\;. \label{bspleq3}
\end{equation}
The quantity $R(x)$ is called the {\it residual}; it is a measure of the accuracy
of the lattice representation.

\subsection{Basis-Spline collocation method}

In the collocation method, one minimizes the residual locally, i.e. one
introduces a set of collocation (data) points $x_\alpha$ and requires
that the residual vanish exactly at these points
\begin{equation}
R(x_\alpha) = 0 \ . \label{bspleq4}
\end{equation}
We multiply Eq.(\ref{bspleq3}) from the left with $\delta(x-x_\alpha)$ and
integrate over x, including a volume element weight function $v(x)$ in
the integrals to emphasize that the formalism applies to arbitrary
curvilinear coordinates. Most cases of interest are covered by a
function of the form
\begin{equation}
v(x) = x^p\;\;\;\;p=\left\{\begin{array}{ll}
                           0\;\;\;&\mbox{Cartesian}\\
                           1\;\;\;&\mbox{Polar}\\
                           2\;\;\;&\mbox{Spherical}
                           \end{array} \right.\;.   \label{bspleq5}
\end{equation}
Using the collocation condition Eq.(\ref{bspleq4}) one obtains
\begin{equation}
[{\cal{O}} f(x)]_{x_\alpha} - g(x_\alpha) = 0 \ . \label{bspleq6}
\end{equation}
Inserting the B-Spline expansion (\ref{bspleq2}) of the function
$f(x)$ this results in
\begin{equation}
\sum_i [{\cal{O}} B]_{\alpha i} a^i - g(x_\alpha) = 0 \ , \label{bspleq7}
\end{equation}
where we have introduced the shorthand notation
\begin{equation}
[{\cal{O}} B]_{\alpha i} = [{\cal{O}} B_i(x)]_{x_\alpha}\; .  \label{bspleq8}
\end{equation}
We eliminate the expansion coefficients $a^i$ in Eq.(\ref{bspleq7})
by introducing the function values at the lattice support
points $x_\alpha$ including both interior and boundary points
\begin{equation}
f_\alpha = f(x_\alpha) = \sum_i B_i(x_\alpha) a^i = \sum_i B_{\alpha i} a^i \; ,
\label{bspleq9}
\end{equation}
\begin{equation}
g_\alpha = g(x_\alpha) = \sum_i B_i(x_\alpha) b^i = \sum_i B_{\alpha i} b^i \; .
\label{bspleq10}
\end{equation}
The matrix $B_{\alpha i}$ is, in general, rectangular. However, it can be made into a
square matrix by adding either periodic
or fixed boundary conditions; this is described in detail in references
\cite{Uma91,WO95,KO96}. In what follows, the new (square) B-Spline
matrix with the boundary conditions added is also denoted by $B_{\alpha i}$,
for simplicity of notation. Because the matrix $B$ is now a square matrix
it can be inverted to eliminate the expansion coefficients $a^i,\ b^i$
\begin{equation}
a^i = \sum_\alpha B^{i \alpha} f_\alpha \;\;\;,\;\;\;
b^i = \sum_\alpha B^{i \alpha} g_\alpha   \label{bspleq11}\; ,
\end{equation}
resulting in
\begin{equation}
\sum_\beta \sum_i [{\cal{O}} B]_{\alpha i} B^{i \beta} f_\beta = g_\alpha \; .
\label{bspleq12}
\end{equation}
Defining the collocation lattice representation of the operator ${\cal{O}}$ via
\begin{equation}
{\cal{O}}_\alpha^\beta = \sum_i [{\cal{O}} B]_{\alpha i} B^{i \beta} \; ,
\label{bspleq13}
\end{equation}
we arrive at the desired lattice representation
\begin{equation}
\sum_\beta {\cal{O}}_\alpha^\beta f_\beta = g_\alpha \ .
\label{bspleq14}
\end{equation}

\subsection{Basis-Spline Galerkin method}

To derive the Galerkin representation, we multiply Eq.(\ref{bspleq3})
from the left with the spline function $B_k(x)$ and integrate over $x$
\begin{eqnarray}
\label{bspleq15}
& &\int v(x) dx B_k(x) {\cal{O}} f(x) - \int v(x) dx B_k(x) g(x) = \nonumber \\
& & \int v(x) dx B_k(x) R(x)\;.
\end{eqnarray}
Various schemes exist to minimize the residual function $R(x)$; in the Galerkin
method one requires that there be no overlap between the
residual and an arbitrary B-spline function
\begin{equation}
\label{bspleq16}
\int v(x) dx B_k(x) R(x) \stackrel{!} = 0 \;.
\end{equation}
This so called {\it Galerkin condition} amounts to a {\it global
reduction of the residual}. We apply the Galerkin condition to
Eq.(\ref{bspleq15}) and insert the B-Spline expansions for $f(x)$
and $g(x)$, Eq.(\ref{bspleq2}), resulting in
\begin{eqnarray}
\sum_i \left[\int v(x) dx B_k(x) {\cal{O}} B_i(x) \right] a^i &-& \nonumber \\
\sum_i \left[\int v(x) dx B_k(x) B_i(x) \right] b^i &=& 0\;.
\label{bspleq17}
\end{eqnarray}
Defining the matrix elements
\begin{eqnarray}
{\cal{O}}_{k i} &=& \int v(x) dx B_k(x) {\cal{O}} B_i(x)\;, \\
G_{k i} &=& \int v(x) dx B_k(x) B_i(x)\;\;.  \label{bspleq18}
\end{eqnarray}
transforms the (differential) operator equation into a matrix $\times$ vector
equation
\begin{equation}
\label{bspleq19}
\sum_i {\cal{O}}_{k i} a^i = \sum_i G_{k i} b^i\;.
\end{equation}
which can be implemented on modern vector or parallel computers with
high efficiency. The matrix $G_{k i}$ is sometimes referred to as the
{\em Gram} matrix; it represents the non vanishing overlap integrals
between different B-Spline functions. This matrix
possesses several highly desirable numerical properties; it is
symmetric, banded, positive definite, and invertible for any
reasonable placement of the knots and any spline order
\begin{equation}
\sum_{k} G^{j k}  G_{k i} = \delta^{j}_{i}\;.
\end{equation}
We use again the relations (\ref{bspleq11}) to eliminate the expansion
coefficients $a^i$ and $b^i$ in Eq.(\ref{bspleq19}) which results in
the matrix equation on the Galerkin lattice
\begin{equation}
\sum_\beta {\cal{O}}_\alpha^\beta f_\beta = g_\alpha \; ,
\end{equation}
with the differential operator definition on the Galerkin lattice
\begin{equation}
{\cal{O}}_\alpha^\beta = \sum_{ijk} B_{\alpha i} G^{ij} {\cal{O}}_{jk} B^{k \beta} \; .
\end{equation}

We are also extending our previous B-spline work to include nonlinear
grids. Use of a nonlinear lattice should be most useful for loosely
bound systems near the proton or neutron drip lines. Non-Cartesian
coordinates necessitate the use of fixed endpoint boundary conditions;
much effort has been directed toward improving the treatment of these
boundaries.

\subsection{2-D lattice representation of HFB wavefunctions and Hamiltonian}
For a given angular momentum projection quantum number $\Omega$, we solve the
eigenvalue problem on a 2-D grid $(r_\alpha,z_\beta)$ where $\alpha = 1,...,m$
and $\beta = 1,...,n$.

The four components of the spinor wavefunction $\psi^{(4)}(r,z)$ are represented on
the 2-D lattice by a product of Basis Spline functions $B_i (x)$ evaluated at
the lattice support points. For example, the spin-up component of the wavefunction
$\phi_2$ is represented in the form
\begin{equation}
\phi_2^{\Omega} (r_\alpha, z_\beta, \uparrow) = \sum_{i,j} B_i^M (r_\alpha) \ B_j^M (z_\beta)
   \ U_2^{ij} \ .
\end{equation}
Hence, the four-dimensional spinor wavefunction in coordinate space $\psi^{(4)}(r,z)$
becomes an array $\psi(N)$ of length $N = 4 \cdot m \cdot n$.

For the lattice representation of the HFB Hamiltonian, we use a hybrid method in which
derivative operators are constructed using the Galerkin method; this
amounts to a global error reduction. Local potentials are represented by the B-Spline collocation method (local error reduction).

The lattice representation transforms the differential operator
equation into a matrix $\times$ vector equation
\begin{equation}
\sum_{\nu=1}^N {\cal{H}}_{\mu}^{\ \nu} \psi^{\Omega}_{\nu} =
            E^{\Omega}_{\mu} \psi^{\Omega}_{\mu} \ \ \ (\mu=1,...,N) \ .
\end{equation}
The 2-D HFB code is written in Fortran 95 and makes extensive use of new
data concepts, dynamic memory allocation and pointer variables. \\

\subsection{2-D B-Spline Poisson Solver}
In the current version of our  HFB-2D-LATTICE code we have solved
the Poisson equation in cylindrical coordinates $(\rho,z)$
\begin{displaymath}
\frac{\partial^2 \phi(\rho,z)}{\partial \rho^2}+
\frac{\phi(\rho,z)}{4\rho^2}+
\frac{\partial^2 \phi(\rho,z)}{\partial z^2}=
-4 \pi e^2 \sqrt{\rho} \ \rho_p(\rho,z)\;,
\end{displaymath}
by applying a large distance expansion of the Coulomb potential $(|{\bf r}| \gg |{\bf r}'| )$
\begin{equation}
\phi_c({\bf r})=e\int{\frac{\rho_p({\bf r}')}{|{\bf r}-{\bf r}'|} d^3 {\bf r}'},
\end{equation}
\begin{eqnarray*}
\frac{1}{|{\bf r}-{\bf r}'|}&=&e^{-{\bf r}' \cdot \vec{\bigtriangledown}}
\frac{1}{|{\bf r}|}=\frac{1}{r}-{\bf r}'\cdot \vec{\bigtriangledown} \frac{1}{r}+
\frac{1}{2} \left( {\bf r}' \cdot \vec{\bigtriangledown} \right)^2 \frac{1}{r}\\ \nonumber
&-&\frac{1}{6} \left( {\bf r}' \cdot \vec{\bigtriangledown} \right)^3 \frac{1}{r}+
\frac{1}{24} \left( {\bf r}' \cdot \vec{\bigtriangledown} \right)^4 \frac{1}{r}+\ldots,\nonumber
\end{eqnarray*}
with the following boundary conditions
\begin{eqnarray}
 \phi_c(\rho \rightarrow  0,z)&\rightarrow& 0\; , \nonumber \\
 \phi_c(\rho,z \rightarrow z_{max})& \rightarrow& 0\; .
\end{eqnarray}
The differential operators are represented on the lattice
using the B-Spline Galerkin method.

\begin{acknowledgments}
This work has been supported by the U.S. Department of Energy under grant
No. DE-FG02-96ER40963 with Vanderbilt University, by the U.S.\
Department of Energy under Contract Nos.\ DE-FG02-96ER40963
(University of Tennessee), DE-AC05-00OR22725 with UT-Battelle, LLC
(Oak Ridge National Laboratory), and DE-FG05-87ER40361 (Joint
Institute for Heavy Ion Research);  by the National Nuclear Security
Administration under the Stewardship Science Academic Alliances
program through DOE Research Grant DE-FG03-03NA00083. Some of the numerical
calculations were carried out at the IBM-RS/6000 SP supercomputer of
the National Energy Research Scientific Computing Center which is
supported by the Office of Science of the U.S. Department of Energy.
Additional computations were performed at Vanderbilt University's ACCRE
multiprocessor platform.

\end{acknowledgments}


\begin{thebibliography}{99}

\bibitem{NNigata}
               J. Dobaczewski, N. Michel, W. Nazarewicz, M. Ploszajczak, and M.V. Stoitsov, in
               ``A New Era of Nuclear Structure Physics'', ed. Y. Suzuki, S. Ohya,
               M. Matsuo \& T. Ohtsubo, World Scientific (2004), pp.162-171.
               
\bibitem{TOU03} E. Ter\'an, V.E. Oberacker, and A.S. Umar, Phys. Rev. C {\bf 67},
                064314 (2003).

\bibitem{OU03} V.E. Oberacker, A.S. Umar, E. Ter\'an, and A. Blazkiewicz,
               Phys. Rev. C {\bf 68}, 064302 (2003).

\bibitem{DN96} J. Dobaczewski, W. Nazarewicz, T.R. Werner, J.F. Berger,
               C.R. Chinn, and J. Decharg\'e, Phys. Rev. {\bf C 53}, 2809 (1996).

\bibitem{OU04} V.E. Oberacker, A.S. Umar, A. Blazkiewicz, and E. Ter\'an,
               in ``A New Era of Nuclear Structure Physics'', ed. Y. Suzuki, S. Ohya,
               M. Matsuo \& T. Ohtsubo, World Scientific (2004), pp.179-183.
               
\bibitem{egidorobledo} J.L. Egido, L. M. Robledo, and Y. Sun, Nucl. Phys. {\bf A560},
               253 (1993).
	       
\bibitem{Stoitsov:2003pd}
                M.~V.~Stoitsov, J.~Dobaczewski, W.~Nazarewicz, S.~Pittel, and D.~J.~Dean,
                Phys.\ Rev.\ C {\bf 68}, 054312 (2003).
                
\bibitem{yamagami} M. Yamagami, K. Matsuyanagi, and M. Matsuo, Nucl. Phys.
                   {\bf A693}, 579 (2001).
                   
\bibitem{Terasaki:1995du}
               J.~Terasaki, P.~H.~Heenen, H.~Flocard, and P.~Bonche,
               Nucl.\ Phys.\ A {\bf 600}, 371 (1996).
	       
\bibitem{Tajima2003}
               N.~Tajima, Phys.\ Rev.\ C {\bf 69}, 034305 (2004).

\bibitem{Ca02} P. Campbell {\it et al.}, Phys. Rev. Lett. {\bf 89}, 082501 (2002).

\bibitem{R-A04} S. Rinta-Antila, S. Kopecky, V.S. Kolhinen, J. Hakala, J. Huikari,
                A. Jokinen, A. Nieminen, and J. \"{A}yst\"{o}, Phys. Rev. C {\bf 70},
                011301(R), (2004).
                
\bibitem{CB98} E. Chabanat, P. Bonche, P. Haensel, J. Meyer, and R. Schaeffer,
               Nucl. Phys. {\bf A635}, 231 (1998); Nucl. Phys. {\bf A643}, 441 (1998).

\bibitem{HR03} {\it Half lives of isomeric states from SF of $^{252}$Cf and large deformations
        in $^{104}$Zr and $^{158}$Sm}, J.K. Hwang, A.V. Ramayya, J.H. Hamilton,
        D. Fong, C.J. Beyer, P.M. Gore, E.F. Jones, E. Ter\'an, V.E. Oberacker, A.S. Umar,
        Y.X. Luo, J.O. Rasmussen, S.J. Zhu, S.C. Wu, I.Y. Lee, P. Fallon, M.A. Stoyer,
        S. J. Asztalos, T.N. Ginter, J.D. Cole, G.M. Ter-Akopian, and R. Donangelo,
        (in preparation).

\bibitem{RD99} P.-G. Reinhard, D.J. Dean, W. Nazarewicz, J. Dobaczewski, J.
               A. Maruhn, and M.R. Strayer, Phys. Rev. C {\bf 60}, 014316 (1999).

\bibitem{SD00} M.V. Stoitsov, J. Dobaczewski, P. Ring, and S. Pittel,
                Phys. Rev. C {\bf 61}, 034311 (2000).

\bibitem{Samyn:2004bm}
               M.~Samyn, S.~Goriely, M.~Bender, and J.~M.~Pearson,
               Phys.\ Rev.\ C {\bf 70}, 044309 (2004).
	       

\bibitem{Ri96} P. Ring, Prog. Part. Nucl. Phys. {\bf 37}, 193 (1996).

\bibitem{LR99} G.A. Lalazissis, S. Raman, and P. Ring,
               Atomic Data and Nuclear Data Tables {\bf 71}, 1 (1999).

\bibitem{AW95} G. Audi and A.H. Wapstra, Nucl. Phys. {\bf A595}, 409 (1995); and
               ``Table of Nuclides'', Brookhaven Nat. Lab.,\url{http://www2.bnl.gov/ton/index.html}.
	       
\bibitem{audi2003} G. Audi, A. H. Wapstra and C. Thibault,
               Nucl. Phys. A {\bf729}, 337-676 (2003).

\bibitem{DeB78} C. De Boor, {\it Practical Guide to Splines}, (Springer Verlag, New York, 1978),
                and references therein.

\bibitem{Uma91} A.S. Umar, J. Wu, M.R. Strayer, and C. Bottcher,
                J. Comp. Phys. {\bf 93}, 426, (1991).

\bibitem{Uma91a} A.S. Umar, M.R. Strayer, J.-S. Wu, D.J. Dean, and M.C. G\"u\c cl\"u, Phys.
                 Rev. C {\bf 44}, 2512 (1991).

\bibitem{WO95} J.C. Wells, V.E. Oberacker, M.R. Strayer and A.S. Umar,
                Int. J. Mod. Phys. C {\bf 6}, 143, (1995).

\bibitem{KO96} D.R. Kegley, V.E. Oberacker, M.R. Strayer, A.S. Umar, and
               J.C. Wells, J. Comp. Phys. {\bf 128}, 197, (1996).

\end{thebibliography}


\end{document}